\documentclass{article}
\pdfoutput=1

 \PassOptionsToPackage{numbers}{natbib}
\usepackage[preprint]{neurips_2019}

\usepackage[utf8]{inputenc} % allow utf-8 input
\usepackage[T1]{fontenc}    % use 8-bit T1 fonts
\usepackage{hyperref}       % hyperlinks
\usepackage{url}            % simple URL typesetting
\usepackage{booktabs}       % professional-quality tables
\usepackage{amsfonts}       % blackboard math symbols
\usepackage{nicefrac}       % compact symbols for 1/2, etc.
\usepackage{microtype}      % microtypography

\usepackage{subcaption}
\usepackage{tikz}
\usetikzlibrary{trees}
\usepackage{multirow}
\usepackage{xcolor}

\usepackage{mdframed}
\usepackage{textcomp}   % for straight single-quote in lstset using upquote=true

\newenvironment{example}[1][]{%
\mdfsetup{frametitle={#1},
linecolor=blue!20,,linewidth=1.2pt,%
frametitlerule=true,%
frametitlebackgroundcolor=gray!20,
innertopmargin=\topskip,}
\begin{mdframed}[]\relax%
}{\end{mdframed}}

\usepackage{listings}
\definecolor{mygray}{rgb}{0.4,0.4,0.4}
\newcommand\myciteauthoryear[1]{\citeauthor{#1}~\cite{#1}}
\newcommand{\tool}{our technique}
\newcommand{\Tool}{Our technique}
\newcommand{\ptool}{the proposed technique}

\usepackage{amsmath}
\newcommand{\esm}[1]{\ensuremath{#1}}
\newcommand{\ms}[1]{\esm{\mathsf{#1}}}
\newcommand\reals{\ms{R}}

\newcommand\xbase{x'}
\newcommand\integratedgrads{\ms{IntegratedGrads}}
\newcommand\sparam{\alpha}
\newcommand\synteq{=}

\title{Deep Learning for Bug-Localization in Student Programs}

\author{%
	Rahul Gupta$^{1}$ \qquad Aditya Kanade$^{1,2}$ \qquad Shirish Shevade$^{1}$ \\
	$^{1}$Department of Computer Science and Automation, \\
	Indian Institute of Science, Bangalore, KA 560012, India \\
	$^{2}$Google Brain,	CA, USA \\
	\texttt{\{rahulg, kanade, shirish\}@iisc.ac.in}
}

\begin{document}	
\maketitle

\begin{abstract}
Providing feedback is an integral part of teaching. 
Most open online courses on programming make use of automated grading systems to support programming assignments and give real-time feedback. 
These systems usually rely on test results to quantify the programs' functional correctness. 
They return failing tests to the students as feedback. 
However, students may find it difficult to debug their programs if they receive no hints about where the bug is and how to fix it. 
In this work, we present the first deep learning based technique that can localize bugs in a faulty program w.r.t.\ a failing test, without even running the program.
At the heart of our technique is a novel tree convolutional neural network which is trained to predict whether a program passes or fails a given test. 
To localize the bugs, we analyze the trained network using a state-of-the-art neural prediction attribution technique and see which lines of the programs make it predict the test outcomes.
Our experiments show that \ptool{} is generally more accurate than two state-of-the-art program-spectrum based and one syntactic difference based bug-localization baselines.
\end{abstract}

% !TeX spellcheck = en_US
\section{Introduction}
\label{section:intro}

Automated grading systems for student programs both check the functional correctness of assignment submissions and provide real-time feedback to students. 
The feedback helps students learn from their mistakes, allowing them to revise and resubmit their work.
In the current practice, automated grading systems rely on running the submissions against a test suite. 
The failing tests are returned to the students as feedback. However, students may find it difficult to debug their programs if they receive no hints about where the bug is and how to fix it.
Although instructors may inspect the code and manually provide such hints in a traditional classroom setting, doing this in an online course with a large number of students is often infeasible.
Therefore, our aim in this work is to develop an automated technique for generating feedback about the error locations corresponding to the failing tests.
Such a technique benefits both instructors and students by allowing instructors to automatically generate hints for students without giving away the complete solution.

\begin{figure}[t]
	\centering
	\begin{minipage}{0.48\linewidth}
		\begin{tikzpicture}
			\node[anchor=south west,inner sep=0] (image) at (0,0) {\includegraphics[width=0.9\linewidth]{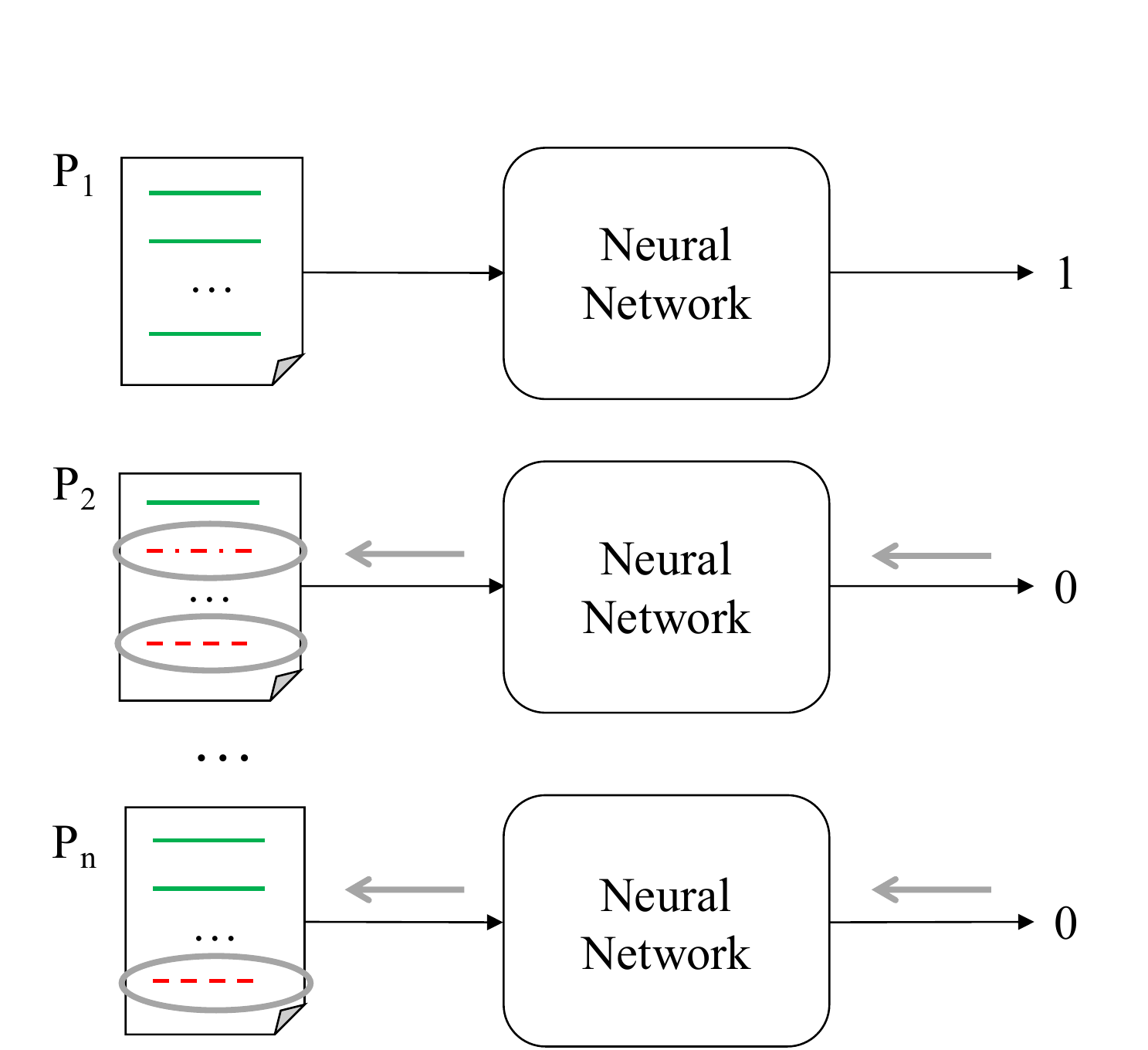}};
			\begin{scope}[x={(image.south east)},y={(image.north west)}]
			\node[] at (0.2,0.97) {\small Input:};
			\node[] at (0.2,0.91) {\small $<$program, test\_id$>$}; 
			\node[] at (0.83,0.97) {\small Output:};
			\node[] at (0.83,0.91) {\small pass:$1$, fail:$0$}; 
			\end{scope}
		\end{tikzpicture}
		\caption{Overview of \tool{}. The buggy lines in the input programs are represented by dashed lines. We omit test ids from the input for brevity. The forward black arrows show the neural network prediction for each input. The thick gray arrows and ovals show the prediction attribution back to the buggy input programs leading to bug-localization.}
		\label{fig:schematic}
	\end{minipage} \hfill
	\begin{minipage}{0.48\linewidth}
		\includegraphics[width=1\linewidth]{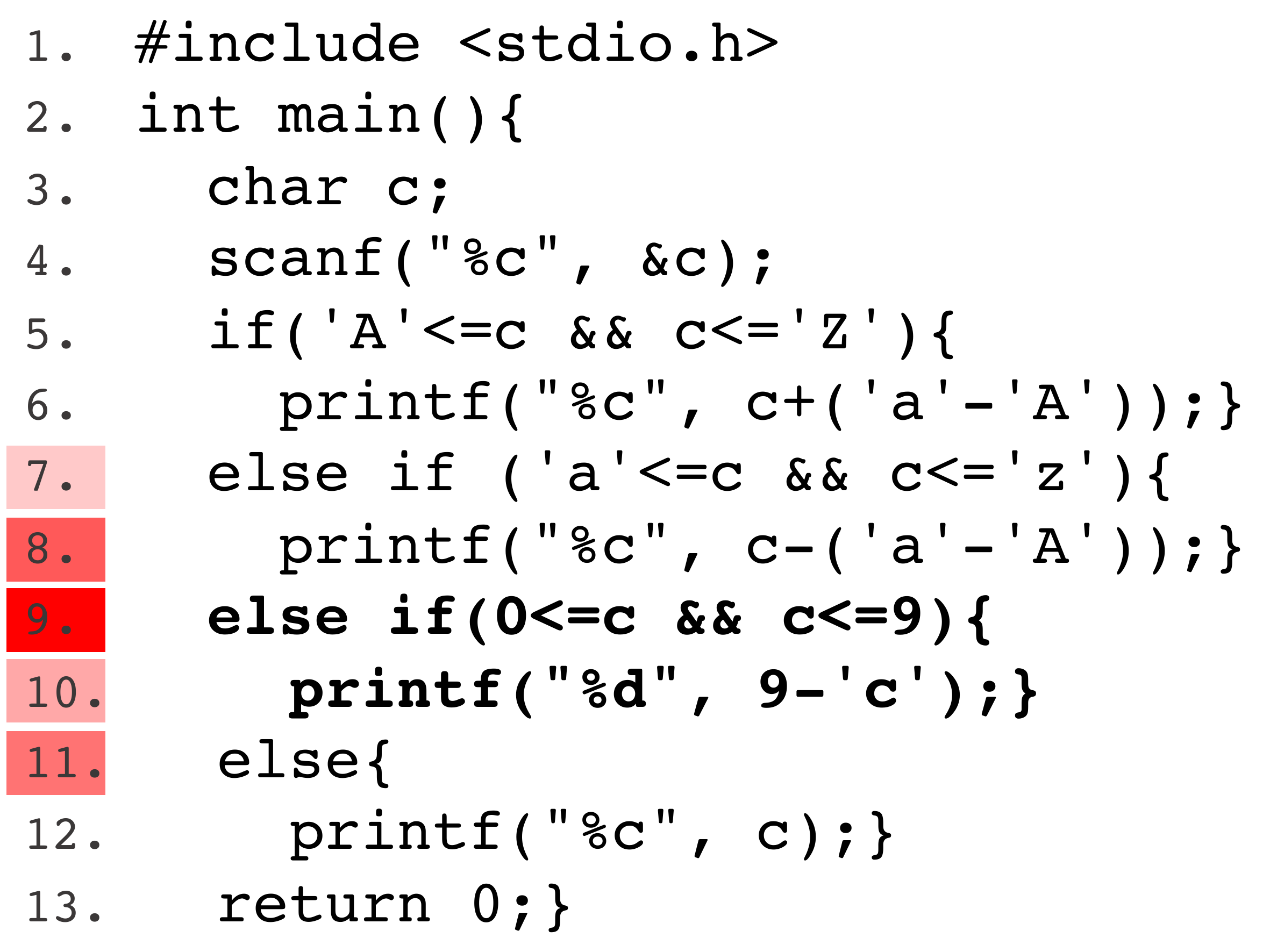}
		\caption{Example to illustrate \tool{}. The program shown has bugs at lines $ 9 $ and $ 10 $ (shown in bold). The top-$ 5 $ suspicious lines returned by \tool{} for this program are marked using a heat-map where darker color indicates higher suspiciousness score.}
		\label{fig:example}
	\end{minipage}
\end{figure}

Towards this, we propose a deep learning based semantic bug-localization technique.
While running a program against a test suite can detect the presence of bugs in the program, locating these bugs requires careful analysis of the program behavior.
Our technique works in two phases. In the first phase, we train a novel tree convolutional neural network to predict whether or not a program passes a given test.
The input to this network is a pair of a program and a test id.
In the second phase, we query a state-of-the-art neural prediction attribution technique~\citep{sundararajan2017axiomatic} to find out which lines of a buggy program make the network predict the failures to localize the bugs.
Figure~\ref{fig:schematic} shows the overview of our technique.
Figure~\ref{fig:example} shows \tool{} in action on a buggy student submission w.r.t.\ a failed test where the test input is $ 0 $ and the expected output is $ 9 $.
For a given character $ c $, the programming task for this submission requires as output the character obtained by reversing its case if $ c $ is an alphabet, or $ 9 - c $ if it is a digit, otherwise c itself. The illustrated submission mishandles the second case in lines $9$ and $10$ and prints the same digit as the input.

Prediction attribution techniques are employed for attributing the prediction of a deep network to its input features. 
For example, for a multi-class image recognition network, a prediction attribution technique can identify the pixels associated with the given class in the given image, and thus can be used for object-localization in spite of being trained on image labels only.
Our work introduces prediction attribution for semantic bug-localization in programs.

Bug-localization is an active field of research in software engineering~\citep{wong2016survey}.
Spectrum-based bug localization approach~\citep{jones2001visualization,ochiai2006} instruments the programs to get the program traces corresponding to both the failing and the passing tests.
In order to locate the bugs in a program, it compares the program statements that are executed in failing test runs against those executed in passing test runs.
While spectrum-based bug-localization exploits correlations between executions of the same program on multiple tests, our technique exploits similarities/differences between the code of multiple programs w.r.t.\ the same test. In this way, the former is a dynamic program analysis approach, whereas the latter is a static program analysis approach.

The existing static approaches for bug-localization in student programs compare a buggy program with a reference implementation~\citep{kaleeswaran2016semi,kim2016apex}.
However, the buggy program and the reference implementation can use different variable names, constant values, and data and control structures, making it extremely difficult to distinguish bug inducing differences from the benign ones.
Doing this requires the use of sophisticated program analysis techniques along with heuristics, which may not work for a different programming language.
In contrast, \tool{} does not require any heuristics and therefore, is programming language agnostic.

Use of machine learning in software engineering research is not new.
Several recent works proposed deep learning based techniques for automated syntactic error repair in student programs~\citep{gupta2017deepfix,bhatia2018neuro,ahmed2018compilation,gupta2019deep}. 
Bugram~\citep{wang2016bugram} is a language model based bug-detection technique.
\myciteauthoryear{pu2016skip} propose a deep learning based technique for both syntactic and semantic error repair in small student programs. Their technique uses a brute-force, enumerative search for detecting and localizing bugs.
Another recent work~\citep{vasic2019neural} proposed a multi-headed LSTM pointer network for joint localization and repair of variable-misuse bugs.
In contrast, ours is a semantic bug-localization technique, which learns to find the location of the buggy statements in a program. 
Unlike these approaches, our technique neither requires explicit bug-localization information for training nor does it perform a brute-force search. 
Instead, it trains a neural network to predict whether or not a program passes a test and analyses gradients of the trained network for bug-localization.
Moreover, our technique is more general and works for all kinds of semantic bugs.
To the best of our knowledge, we are the first to propose a general deep learning technique for semantic bug-localization in programs w.r.t.\ failing tests.

We train and evaluate \tool{} on C programs written by students for $ 29 $ different programming tasks in an introductory programming course. 
The dataset comes with $ 231 $ instructor written tests for these tasks. 
Thus, programs for each task are tested against about $ 8 $ tests on an average.
We compare our technique with three baselines which include two state-of-the-art, program-spectrum based techniques~\cite{jones2001visualization,ochiai2006} and one syntactic difference based technique.
Our experiments demonstrate that \ptool{} is more accurate than them in most cases.
The main contributions of this work are as follows:
\begin{enumerate}
	\item It proposes a novel encoding of program ASTs and a tree convolutional neural network that allow efficient batch training for arbitrarily shaped trees.
	\item It presents the first deep learning based general technique for semantic bug-localization in programs. It also introduces prediction attribution in the context of programs.
	\item The proposed technique is evaluated on thousands of erroneous C programs with encouraging results.
	It successfully localized a wide variety of semantic bugs, including wrong conditionals, assignments, output formatting and memory allocation, among others.
	We provide several concrete examples of these in the appendix.
\end{enumerate}
Both the dataset and the implementation of \tool{} will be open sourced.

% !TeX spellcheck = en_US
\section{Background: Prediction Attribution}
\label{background}
Prediction attribution techniques attribute the prediction of a deep network to its input features.
For our task of bug-localization, we use a state-of-the-art prediction attribution technique called integrated gradients~\citep{sundararajan2017axiomatic}.
This technique has been shown to be effective in domains as diverse as object recognition, medical imaging, question classification, and neural machine translation among others.
In Section~\ref{subsection:attribution}, we explain how we leverage integrated gradients for bug-localization in programs.
Here we describe this technique briefly. 
For more details, we refer our readers to the work of~\myciteauthoryear{sundararajan2017axiomatic}.

When assigning credit for a prediction to a certain feature in the input, the absence of the feature is required as a baseline for comparing outcomes. 
This absence is modeled as a single baseline input on which the prediction of the neural network is ``neutral" i.e., conveys a complete absence of signal.
For example, in object recognition networks, the black image can be considered as a neutral baseline.
Integrated gradients technique distributes the difference between the two outputs (corresponding to the input of interest and the baseline) to the individual input features.

More formally, for a deep network representing a function $F: \reals^n \rightarrow [0,1]$  where input $x \in \reals^n$, and baseline $\xbase \in \reals^n$; integrated gradients are defined as the path integral of the gradients along the straight-line path from the baseline $\xbase$ to the input $x$.
For $x$ and $\xbase$, the integrated gradient along the $i^{th}$ dimension is defined as follows:
\begin{align*}
\integratedgrads_i(x) & \synteq (x_i-\xbase_i)\times\int_{\sparam=0}^{1} \tfrac{\partial F(\xbase + \sparam(x-\xbase))}{\partial x_i  }~d\sparam
\end{align*}

If $F: \reals^n \rightarrow \reals$ is differentiable almost everywhere,
then $$\Sigma_{i=1}^{n} \integratedgrads_i(x) = F(x) - F(\xbase)$$
If the baseline $\xbase$ is chosen in a way such that the prediction at the baseline is near zero ($F(\xbase) \approx 0$), then resulting attributions have an interpretation that ignores the baseline and amounts to distributing the output to the individual input features.
The integrated gradients can be efficiently approximated via summing the gradients at points occurring at sufficiently small intervals along the straight-line path from the baseline $\xbase$ to the input $x$:
\begin{equation*}
\begin{split}
\integratedgrads_i^{approx}(x) \synteq 
(x_i-\xbase_i)\times \Sigma_{k=1}^m  \tfrac{\partial F(\xbase + \tfrac{k}{m}(x-\xbase)))}{\partial x_i}\times\tfrac{1}{m}
\end{split}
\end{equation*}
where $m$ is the number of steps in the Riemman approximation of the
integral of integrated gradients~\citep{sundararajan2017axiomatic}.
% !TeX spellcheck = en_US

\section{Technical Details}

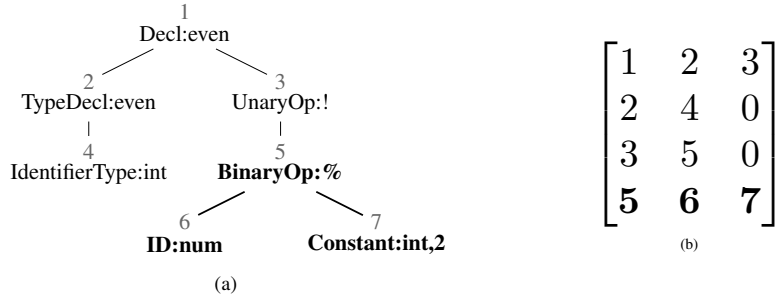
\begin{figure}
	\centering
	\scalebox{0.85}{
	\begin{subfigure}{0.5\linewidth}
		\centering
		\begin{tikzpicture}[level distance=1.1cm,
		level 1/.style={sibling distance=3cm},
		level 2/.style={sibling distance=3cm}]
		
		\node {Decl:even}
		child {
			node [above=3pt] {\color{black!20!gray}{2}}
			node {TypeDecl:even}
			child {
				node [above=3pt] {\color{black!20!gray}{4}}
				node {IdentifierType:int}
				edge from parent [draw=black]
			}
			edge from parent [draw=black]
		}
		child {
			node [above=3pt] {\color{black!20!gray}{3}}
			node {UnaryOp:!}
			child {
				node [above=3pt] {\color{black!20!gray}{5}}
				node {\textbf{BinaryOp:\%}}
				edge from parent [draw=black]
				child {
					node [above=3pt] {\color{black!20!gray}{6}}
					node {\textbf{ID:num}}
					edge from parent [draw=black,thick]	
				}
				child {
					node [above=3pt] {\color{black!20!gray}{7}}
					node {\textbf{Constant:int,2}}
					edge from parent [draw=black,thick]	
				}
			}
			edge from parent [draw=black]
		}
		node [above=3pt] {\color{black!20!gray}{1}};
		\end{tikzpicture}
		\caption{}
		\label{fig:ast}
	\end{subfigure}
	} \qquad
	\scalebox{0.65}{
	\begin{subfigure}{0.5\linewidth}
		\centering

		\begin{tikzpicture}
		\node[anchor=south west,inner sep=0] (image) at (0,0) {\includegraphics[width=0.57\textwidth]{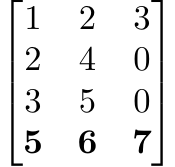}};
		\end{tikzpicture}
		\caption{}
		\label{fig:programEncoding}
	\end{subfigure}
	}
	\caption{\ref{fig:ast}: AST of the code snippet: \texttt{int even=!(num\%2)}. For each node, its visiting order is also shown in the breadth-first traversal of the AST. 
		\ref{fig:programEncoding}: 2-d matrix representation of the AST shown in Figure~\ref{fig:ast}. The matrix shows node positions instead of the  nodes themselves to avoid clutter. For example, the last row corresponds to the highlighted subtree from Figure~\ref{fig:ast}.}
	\label{fig:astAndEncoding}
\end{figure}

We divide our bug-localization approach in two phases. In the first phase, we train a neural network to predict whether or not a program passes the test corresponding to a given test id. 
This is essentially a classification problem with two inputs: program text and a test id, where we have multiple passing and failing programs (which map to different class labels) for each test id. Though different programs are used at test time, they share test ids with the training examples.
In the second phase, we perform bug-localization by identifying the patterns that help the neural network in correct classification.
Note that the neural network is only given the test id along with the program as input. 
It is not provided with the actual inputs and the expected outputs of the tests as it does not know how to execute the program.
The learning is based only on the syntactic patterns present or absent in the programs.

\subsection{Phase $ 1 $: Tree Convolutional Neural Network for Test Success/Failure Prediction}
Use of machine learning in software engineering research is not new.
Many works exist which use machine learning algorithms on programs for different tasks such as code completion, automated repair of syntactic errors, and program synthesis among others. 
Details of many such techniques can be found in a survey~\citep{allamanis2018survey} and the references therein.
Most of these works use recurrent neural networks~(RNNs)~\citep{gupta2017deepfix,vasic2019neural} and convolutional neural networks~(CNNs)~\citep{mou2016convolutional}.
Our initial experiments with multiple variants of both RNNs and CNNs suggested the latter to be better suited for our task.
CNNs are designed to capture spatial neighborhood information in data and are generally used with inputs having grid-like structure such as images~\citep{goodfellow2016deep}.
On their own, they may fail to capture the hierarchical structures present in programs.
To address this,~\myciteauthoryear{mou2016convolutional} proposed tree based CNNs.
However, the design of their custom filter is difficult to implement and train as it does not allow batch computation over variable-sized programs and trees.
Therefore, we propose a novel tree convolutional network which uses specialized program encoding and convolution filters to capture the tree structural information present in programs, allowing us to not only batch variable-sized programs but also use the well optimized CNN implementations provided by the popular deep learning frameworks of the day.

\subsubsection{Program Encoding}
Programs have rich structural information, which is explicitly represented by their abstract syntax trees~(ASTs).
Figure~\ref{fig:ast} shows the AST of the following code snippet: \\
\centerline{
\texttt{int even=!(num \% 2);}
}
Each node in an AST represents an abstract construct in the program source code.
We encode programs in such a way that their explicit tree structural information is captured by CNNs easily. To do this, we convert the AST of a program into an adjacency list like representation as follows.
First, we flatten the tree in the breadth-first traversal. In the second step, each non-terminal node in this flattened tree is replaced by a list with the first element in the list being the node itself and the rest of the elements are its direct children nodes ordered from left to right. As terminal nodes do not hold any structure by themselves, we discard them at this step.

Next, we convert this representation into a 2-dimensional matrix for feeding it to a CNN.
We do that by padding subtrees with dummy nodes to make them of equal size across all programs in our dataset.
We also pad the programs with dummy subtrees to make each program have the same number of subtrees.
This way each program is encoded into a 2-dimensional matrix of size $max\_subtrees \times max\_nodes$, where $max\_subtrees$ and $max\_nodes$ denote the maximum number of subtrees and the maximum number of nodes in a depth-$1$ subtree across all programs in our dataset, respectively.
Figure~\ref{fig:programEncoding} shows the 2-dimensional matrix representation for the AST shown in Figure~\ref{fig:ast} where $ 0 $ indicates padding.
In this representation, each row of the encoded matrix corresponds to a depth-$1$ subtree in the program AST.
Moreover, contiguous subsets of rows of an encoded matrix correspond to larger subtrees in the program AST.
Note that this encoding ensures that the tree structural information of a program is captured by the spatial neighborhood of elements within a row of its encoded matrix; allowing us to use CNNs with simple convolution filters which can extract features from complete subtrees at a time and not just from any random subset of nodes.

Next, we create a shared vocabulary across all program ASTs in our dataset. 
The vocabulary retains all the AST nodes such as non-terminals, keywords, and literals except for the identifiers~(variable and function names) without any modification.
Identifiers are included in the vocabulary after normalization. This is done by creating a small set of placeholders and mapping each distinct identifier in a program to a unique placeholder in our set. The size of the placeholder set is kept large enough to allow this normalization for every program in our dataset.
This transformation prevents the identifiers from introducing rarely used tokens in the vocabulary without changing the semantics of the program they appear in.

\subsubsection{Neural Network Architecture}
\label{subsection:tcnn}

\begin{figure}[t]
	\centering
	\includegraphics[width=0.8\textwidth]{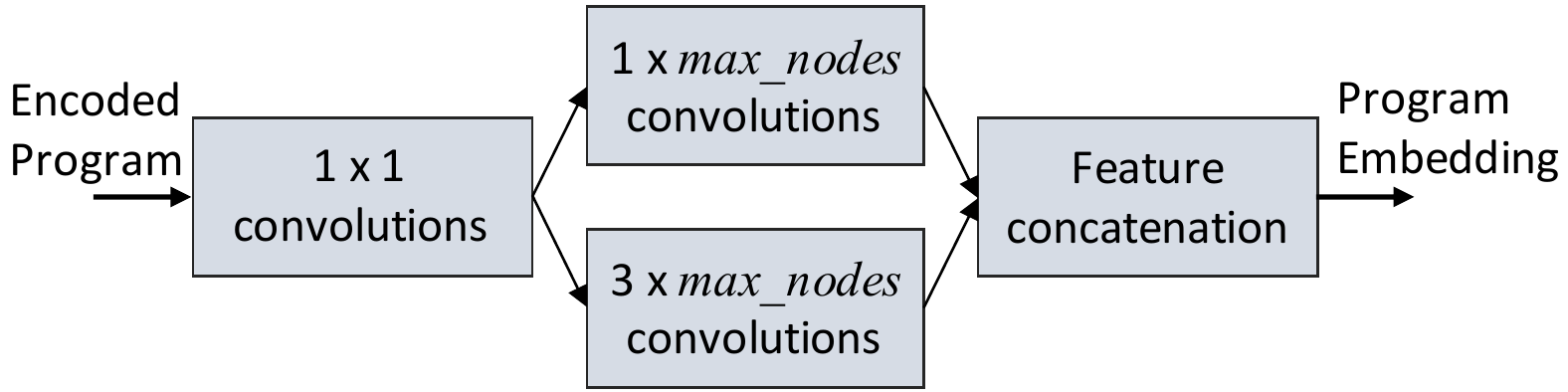}
	\caption{Tree convolution over the encoded program AST input. Variable $ max\_nodes $ represents the maximum number of nodes in a depth-$1$ subtree across all programs in our dataset.}
	\label{fig:TCNN}
\end{figure}

Given a pair of a program and a test id as input, our learning task is to predict the binary test result i.e., failure or success of the input program on the test corresponding to the given test id.
To do this, we first encode the input program into its 2-d matrix representation as discussed above. Each element~(node) of the matrix is then replaced by its index in the shared vocabulary which is then embedded into a $ 24 $-dimensional dense vector using an embedding layer.
The output 3-d matrix is then passed through a convolutional neural network to compute a dense representation of the input program as shown in Figure~\ref{fig:TCNN}.
The first convolutional layer of our network extracts features from a single node at a time.
The output is then passed through two independent convolutional layers.
The first of these two layers applies filters overlapping one whole row at a time with a stride of one row. The filter for second overlaps three rows at a time with a stride of three rows.
As discussed earlier, each row of the program encoding matrix represents a depth-$1$ subtree of the program AST.
This makes the last two convolutional layers detect features for one subtree and three subtrees at a time.

The resulting features from both these layers are then concatenated to get the program embedding.
Next, we embed the test id into a $ 5 $-dimensional dense vector using another embedding layer.
It is then concatenated with the program embedding and passed through three fully connected non-linear layers to generate the binary result prediction.
We call our model tree convolutional neural network~(TCNN).

\subsection{Phase 2: Prediction Attribution for Bug-Localization}
\label{subsection:attribution}
For a pair of a buggy program and a test id, such that the program fails the test, our aim is to localize the buggy line(s) in the program that are responsible for this failure.
If our trained model predicts the failure for such a pair correctly, then we can query a prediction attribution technique to assign the blame for the prediction to the input program features to find these buggy line(s).
As discussed earlier, we use the integrated gradients~\citep{sundararajan2017axiomatic} technique for this purpose.

In order to attribute the prediction of a network to its input, this technique requires a neutral baseline which conveys the complete absence of signal.
\citeauthor{sundararajan2017axiomatic} suggest black images for object recognition networks and all-zero input embedding vectors for text based networks as baselines.
However, using an all-zero input embedding vector as a baseline for all the buggy programs does not work for our task.
Instead, we propose to use a correct program similar\footnote{denotes both syntactic and semantic similarity.} to the input buggy program as a baseline for attribution.
This works because the correct program does not have any patterns which cause bugs and hence, conveys the complete absence of the signal required for the network to predict the failure.
Furthermore, we are interested in capturing only those changes which introduce bugs in the input program and not the benign changes which do not introduce any bugs.
This justifies the use of a similar program as a baseline.
Using a very different correct program as a baseline would unnecessarily distribute the output difference to benign changes which would lead to the undesirable outcome of localizing them as bugs.

For a buggy submission by a student, we find the baseline from the set of correct submissions by other students, as follows. 
First, we compute the embeddings of all the correct programs using our tree CNN. Then we compute the cosine distance of these embeddings from the buggy program embedding. The correct program with the minimum cosine distance is used as the baseline for attribution. 

The integrated gradient technique assigns credit values to each element of an embedded AST node, 
which are averaged to get the credit value for that node. 
As bug-localization techniques usually localize bugs to the program lines, we further average the credit values for nodes to get the credit values for each line in the program. The nodes corresponding to a line are identified using the parser used to generate the ASTs.
We interpret the credit value for a line as the suspiciousness scores for that line to be buggy.
Finally, we return a ranked list of program lines sorted in decreasing order of their suspiciousness scores.
% !TeX spellcheck = en_US
\section{Experiments}

\subsection{Dataset}

\begin{table*}[t]
	\centering
	\caption{Dataset statistics.}  
	\begin{tabular}{rrrrrr}
		\toprule
		\multicolumn{2}{c}{Avg.\ programs} & \multicolumn{1}{c}{Avg.} & \multicolumn{1}{c}{Avg.}  & \multicolumn{1}{c}{Avg.} \\
		\multicolumn{2}{c}{per task} & \multicolumn{1}{c}{tests} & \multicolumn{1}{c}{submissions} & \multicolumn{1}{c}{lines per} \\
		\multicolumn{1}{c}{Correct} & \multicolumn{1}{c}{Buggy} & \multicolumn{1}{c}{per task} & \multicolumn{1}{c}{per student} & \multicolumn{1}{c}{submission} \\
		\midrule
		$ 350 $		&   $ 1007 $	&  $ 8 $ 	&  $ 18 $  &   $ 25 $ \\
		\bottomrule
	\end{tabular}
	\label{tab:dataset}
\end{table*}

For training and evaluation, we use student written C programs for $ 29 $ different programming tasks in an introductory programming course. 
The problem statements of these tasks are quite diverse requiring students to implement concepts such as simple integer arithmetic, array and string operations, backtracking, and dynamic programming. Solving these require various language constructs such as scalar and multi-dimensional arrays variables, conditionals, nested loops, recursion and functions.
We list the problem statements for some of the programming tasks in the appendix.
Our dataset comes with the instructor provided test suite for each programming task. The dataset contains a total of $231$ tests across these $ 29 $ programming tasks.
Note that we work only with the tests written by the instructors of this course and do not write or generate any additional tests. 
A program is tested only against tests from the same programming task it is written for. This is assumed in the discussion henceforth. Each program in our dataset contains about $ 25 $ lines of code on average.

Table~\ref{tab:dataset} shows the dataset statistics. 
We have two classes of programs in our dataset, (1) programs which pass all the tests (henceforth, correct programs), and (2) programs which fail and pass at least one test each~(henceforth, buggy programs). 
We observed that programs which do not pass even a single test to be almost entirely incorrect. Such program do not benefit from bug-localization and hence we discard them.
Now for each test, we take maximum $ 700 $ programs that pass it~(including buggy programs that fail on other tests) and maximum $ 700 $ programs that fail it.
Next, we generate subtrees for each of these programs using pycparser~\cite{pycparser}.
In order to remove unusually bigger programs, we discard the last one percentile of these programs arranged in the increasing order of their size. 
Across all the remaining programs, $ max\_nodes $ and $ max\_subtrees $ come out to be $ 21 $ and $ 149 $, respectively.
Pairing these programs with their corresponding test ids results in a dataset with around $270k$ examples.
We set aside $5\%$ of this dataset for validation and use the rest for training.

\paragraph{Evaluation Dataset}
Evaluating bug-localization accuracy on a program requires the ground truth in the form of bug locations in that program.
As the programs in our dataset come without the ground truth, we try to find that automatically by comparing the buggy programs to their corrected versions.
For this, we use Python's \texttt{difflib} to find line-wise differences, the `diff', between a buggy and a correct program.
We do this for every pair of buggy and correct programs that are solutions to the same programming task and are written by the same student.
Note that this is only done to find the ground truth for evaluation. Our technique does not use the corrected version of an incorrect program written by the same student.
We include a buggy program in our evaluation set only if we can find a correct program with which its diff is smaller than five lines.
This gives us $ 2136 $ buggy programs containing $ 3022 $ buggy lines in our evaluation set.  
Pairing these programs with their corresponding failing test ids results in  $ 7557 $ pairs.
We ensure that these pairs do not overlap with the training data.

In order to identify the buggy lines from the diff, we first categorize each patch appearing in the diff into three categories: (1) insertion of correct line(s), (2) deletion of buggy line(s), and (3) replacement of buggy line(s) with correct line(s).
Next, we mark all the lines appearing in the deletion and replacement categories as buggy.
For the lines in the first category, we mark their preceding line as buggy.
For a program with a single buggy line, it is obvious that all the failing tests are caused by that line.
However, for the programs with multiple buggy lines, we need to figure out the buggy line(s) corresponding to each failing test.
We do this as follows.

For a buggy program and its diff with the correct implementation, first we create all possible partially corrected versions of the buggy program by applying all non-trivial subsets of the diff generated patches.
Next, we run partially corrected program versions against the test suite and for each program, mark the buggy line(s) excluded from the partial fix as the potential cause for all the tests that the program fails.
Next, we go over these partially fixed programs in the increasing order of the number of buggy lines they have. For each program we mark the buggy lines in that program as a cause for a failing test if a program having a subset of buggy lines does not also fail that test.
This procedure is similar in spirit to delta debugging approach~\citep{Zeller1999Yesterday}, which uses unit tests to narrow down bug causing lines while removing lines that are not responsible for reproducing the bug.

\subsection{Training}
We implement our technique in Keras~\citep{chollet2015keras} using Tensorflow~\citep{abadi2016tensorflow} as back-end.
We find a suitable configuration of the tree convolutional neural network through experimentation. Our vocabulary has $ 1213 $ tokens after identifier-name normalization. We train our model using the ADAM optimizer~\citep{kingma2015adam} with a learning rate of $ 0.0001 $. We train our model for $ 50 $ epochs, which takes about $ 1 $ hour on an Intel(R) Xeon(R) Gold 6126 machine, clocked at 2.60GHz with $ 64 $GB of RAM and equipped with an NVIDIA Tesla P100 GPU accelerator.
Our model achieves the training and validation accuracies of $ 99.9\% $ and $ 96\% $, respectively.
We use $ 100 $ steps for approximating the integrated gradient for bug-localization.

\subsection{Evaluation}

\begin{table*}[t]
	\centering
	\caption{Comparison of \ptool{} with three baselines. Top-$k$ denotes the number of buggy lines reported in their decreasing order of suspiciousness score. }
	\scalebox{1}{
		\begin{tabular}{@{}llrrrr@{}}
			\toprule
			\multicolumn{1}{c}{Technique \&} & \multicolumn{1}{c}{Evaluation} & \multicolumn{1}{c}{Localization} & \multicolumn{3}{c}{Bug-localization result} \\ 
			\multicolumn{1}{c}{Configuration} & \multicolumn{1}{c}{metric} & \multicolumn{1}{c}{queries} & \multicolumn{1}{c}{Top-10} & \multicolumn{1}{c}{Top-5} & \multicolumn{1}{c}{Top-1} 
			\\ \midrule
			\multirow{3}{*}{\shortstack[l]{Proposed\\technique}} & $\langle P, t \rangle$ pairs & $ 4117 $ & $ 3134 $ ($ 76.12\% $) & $ 2032 $ ($ 49.36\% $) & $ 561 $ ($ 13.63\% $) 
			\\
			& Lines & $ 2071 $ & $ 1518 $ ($ 73.30\% $) & $ 1020 $ ($ 49.25\% $) & $ 301 $ ($ 14.53\% $) 
			\\
			& Programs & $ 1449 $ & $ \textbf{1164} $ ($ 80.33\% $) & $ 833 $ ($ 57.49\% $) & $ 294 $ ($ 20.29\% $) 
			\\ [2 pt]
			Tarantula-1 & \multirow{2}{*}{Programs} & \multirow{2}{*}{$ 1449 $} & $ 964 $ ($ 66.53\% $) & $ 456 $ ($ 31.47\% $) & $ 6 $ ($ 0.41\% $) 
			\\
			Ochiai-1 &  &  & $ 1130 $ ($ 77.98\% $) & $ 796 $ ($ 54.93\% $) & $ 227 $ ($ 15.67\% $) 
			\\  [2 pt]
			Tarantula-* & \multirow{2}{*}{Programs} & \multirow{2}{*}{$ 1449 $} & $ 1141 $ ($ 78.74\% $) & $ 791 $ ($ 54.59\% $) & $ 311 $ ($ 21.46\% $) 
			\\
			Ochiai-* &  & & $ 1151 $ ($ 79.43\% $) & $ \textbf{835} $ ($ 57.63\% $) & $ \textbf{385} $ ($ 26.57\% $) 
			\\  [2 pt]
			Diff-based & Programs & $ 1449 $ & $ 623 $ ($ 43.00\% $) & $ 122 $ ($ 8.42\%$) & $ 0 $ ($ 0.00 $\%)  
			\\ 
			\midrule
			NBL rank & & & ($ 1/6 $) & ($ 2/6 $) & ($ 3/6 $) \\
			\bottomrule
	\end{tabular}}
	\label{tab:results}
\end{table*}

In the first phase, we use the trained model to predict the success/failure of each example pair of a buggy program and a test id, $\langle P, t \rangle $ from the evaluation dataset.
On these pairs, the classification accuracy of the trained model is only $ 54.48\% $.
This is much lower than its validation accuracy of $ 96\% $. The explanation for such a big difference lies in the way the two datasets are constructed.
The pairs in the validation set are chosen randomly from the complete dataset and therefore their distribution is similar to the pairs in the training dataset. Also, both these datasets consist of pairs associated with both success and failure classes.
On the other hand, recall that the evaluation set contains pairs associated only with the failure class. Furthermore, the buggy programs in these pairs are chosen because we could find their corrected versions with a reasonably small syntactic difference between them.
Thus, the relatively lower accuracy of our model on the evaluation set stems from the fact that its distribution is different from that of training and validation sets and is not actually a limitation of the model.
This is also evident from the fact that the test accuracy increases to about $ 72\% $ if the evaluation set includes pairs associated with both success and failure classes instead of just failure class for the the same programs in the evaluation set.

In the second phase, we query the attribution technique for bug-localization of those pairs of programs and tests for which the model prediction in the earlier phase is correct.
We evaluate the bug-localization performance of \tool{} on the following three metrics: ($ 1 $) the number of pairs for which at least one of the lines responsible for the program failing the test is localized, ($ 2 $) the number of programs for which at least one buggy line is localized, and ($ 3 $) the number of buggy lines localized across all programs.
As shown in Table~\ref{tab:results}, out of the $ 1449 $ programs for which the localization query is made, \tool{} is able to localize at least one bug for more than $ 80\% $ of them, when reporting top-$ 10 $ suspicious lines.
It also proved to be effective in bug-localization for programs having multiple bugs. Out of $ 756 $ such programs in the evaluation set, \tool{} localized more than one bug for $ 314 $ programs, when reporting top-$ 10 $ suspicious lines.

\paragraph{Comparison with Baselines} In Table~\ref{tab:results}, we compare \tool{} with three baselines including two state-of-the-art program-spectrum based techniques, namely Tarantula~\citep{jones2001visualization} and Ochiai~\citep{ochiai2006} and one syntactic difference based approach.
This comparison is made only on those pairs in the evaluation set which our model  classifies correctly. The metric used for this comparison is the number of programs for which at least one bug is localized. The other two metrics, namely, number of $\langle P, t \rangle $ pairs and buggy lines localized also yield similar results.

A program-spectrum records which components of a program are covered, and which are not during an execution. Tarantula and Ochiai compare the program-spectra corresponding to all the failing tests to that of all the passing tests. The only difference between them is that they use different formulae to calculate the suspiciousness scores of program statements.
As \tool{} performs bug-localization w.r.t.\ one failing test at a time, we restrict these techniques to use only one failing test at a time for a fair comparison.
We use them in two configurations. In the first, they are restricted to just one passing test, chosen randomly and in the second, they use all the passing tests. These configurations are denoted by suffixing `-$ 1 $' and `-*' to the names of the techniques, respectively.
The syntactic difference based approach is the same as the one described earlier for finding the actual bug locations~(ground truth) for the programs in the evaluation set. The only difference is that now the reference implementation for a buggy program submitted by a student is searched within the set of correct programs submitted by other students. This is done for both this approach and our technique.

It can be seen that \tool{} outperforms both Tarantula-$ 1 $ and Ochiai-$ 1 $ ~(when they use only one passing test) in top-$k$ results for all values of $ k $.
However, with the extra benefit of using all passing tests, they both outperform \tool{} in top-$ 1 $ results.
Nevertheless, even in this scenario, \tool{} outperforms both of them in top-$ 10 $ results. In top-$ 5 $ results, \tool{} outperforms Tarantula-*, while matching the performance of Ochiai-*.
\tool{} also completely outperforms the syntactic difference based technique with a high margin.

\paragraph{Qualitative Evaluation}
In our analysis, we found that \ptool{} localized almost all kinds of bugs appearing in the evaluation set programs. Some of these include wrong assignments, conditions, for-loops, memory allocations, output formatting, incorrectly reading program inputs, and missing code among others.
We provide a number of concrete examples illustrating our bug-localization results in the appendix.

\Tool{} compares a buggy program to a closely similar correct program using neural attribution. This comparison is designed to search for the bug-causing differences in the buggy program while ignoring the benign ones. As our technique is not engineered to target a predefined set of patterns when searching for differences, in principle, it should be able to localize all kinds of bugs. Therefore, we call our technique a general semantic bug-localization technique. 

\subsection{Faster Search for Baseline Programs through Clustering}
As discussed earlier, we calculate the cosine distance between the embeddings of a given buggy program with all correct programs. When the number of correct programs is large, it can be expensive to search through all of them for each buggy program.
To mitigate this, we cluster all the programs first using the K-means clustering algorithm on their embeddings. 
Now for each buggy program, we search for the baseline only within the set of correct programs present in its cluster.
Note that both clustering and search are performed on programs from the same programming task.
We set the number of clusters to $ 5 $. We arrive at this value through experimentation.
Our results show that clustering affects the bug-localization accuracy by less than $ 0.5\% $ in every metric while reducing the cost of baseline search by a factor of $ 5 $.

% !TeX spellcheck = en_US
\section{Related Work}

We discussed spectrum and diff based bug-localization approaches for student programs earlier in Section~\ref{section:intro}.
We also compared our technique with them empirically in the previous section.
In Section~\ref{section:intro}, we gave an overview of the recent developments in learning based software engineering research as well. In this section, we review two more approaches for feedback generation for student programs.

Program repair techniques are extensively used for feedback generation for logical errors in student programs. AutoGrader~\citep{singh2013automated} takes as input a buggy student program, along with a reference solution and a set of potential corrections in the form of expression rewrite rules and searches for a set of minimal corrections using program synthesis. 
Refazer~\citep{rolim2017learning} learns programs transformations from example code edits made by students using a hand designed domain specific language, and then uses these transformations to repair buggy student submissions. 
Unlike these approaches, our approach is completely automatic and requires no inputs from the instructor.
Most program repair techniques first use an off-the-shelf bug-localization technique to get a list of potential buggy statements. On these statements, the actual repair is performed. We believe that our technique can also be fruitfully integrated into such program repair techniques.

Another common approach to feedback generation is program clustering where student submissions having similar features are grouped together in clusters. 
The clusters are typically used in the following two ways:
(1) the feedback is generated manually for a representative program in each cluster and then customized to other members of the cluster automatically~\citep{nguyen2014codewebs,Piech2015,glassman2015overcode}, and (2) for a buggy program, a reference implementation is selected from the same cluster, which is then compared to the buggy program to generate a repair hint~\citep{kaleeswaran2016semi,gulwani2018automated,wang2018search,Sharma2018}.
The clusters are created either using heuristics based on program analysis techniques~\citep{glassman2015overcode,kaleeswaran2016semi,gulwani2018automated,wang2018search,Sharma2018} or using program execution on a set of inputs~\citep{nguyen2014codewebs,Piech2015}. 
Unlike these approaches, we cluster programs using k-means clustering algorithm on the embeddings learned on program ASTs, which does not require any heuristics and therefore, is programming language agnostic.
% !TeX spellcheck = en_US
\section{Conclusions and Future Work}
We present the first deep learning based general technique for semantic bug-localization in student programs w.r.t.\ failing tests.
At the heart of our technique is a novel tree convolution neural network which is trained to predict whether or not a program passes a given test.
Once trained, we use a state-of-the-art neural prediction attribution technique to find out which lines of the programs make the network predict the failures to localize the bugs.
We compared our technique with three baseline including one static and two state-of-the-art dynamic bug-localization techniques.
Our experiments demonstrate that our technique outperforms all three baselines in most of the cases.

We evaluate our technique only on student programs. 
It will be an interesting future work to use it for arbitrary programs in the context of regression testing~\citep{yoo2012regression}, i.e., to localize bugs in a program w.r.t.\ the failing tests which were passing with the earlier version(s) of that program.
Our technique is programming language agnostic and has been evaluated on C programs. 
In future, we will experiment with other programming languages as well.
We also plan to extend this work to achieve neural program repair.
While our bug-localization technique required only a discriminative network, a neural program repair technique would require a generative model to predict the patches for fixing bugs. It will be interesting to see if our tree convolution neural network  can be adapted to do generative modeling of patches as well.

\section*{Acknowledgments}
We thank Sonata Software Ltd.\ for partially funding this work. 
We also thank the anonymous reviewers for their helpful feedback on the first version of the paper.

\bibliographystyle{plainnat}
\bibliography{bibtex}

\clearpage
\appendix
% !TeX spellcheck = en_US
\section{Appendix}

\subsection{Problem Statements for Some of the Programming Tasks}

\vspace{0.5 cm}
\begin{example}
You want to create an intelligent machine which can perform linear algebra for you. In linear algebra, we often encounter identity matrices. Therefore, teaching computers to recognize whether a matrix is identity or not is one of the tasks that you must perform in your quest to build such a machine. In this problem, you'll write a program to check whether a given matrix is identity or not.

In the first line, you'll be given n, which will be the number of rows and number of columns in identity matrix. In the next n lines, you'll be given entries of the matrix with each row in a new line. If the matrix is identity, then print GIVEN n x n matrix is an IDENTITY MATRIX. Otherwise, print GIVEN n x n matrix is NOT an IDENTITY MATRIX. Here, n is the dimension of the matrix.

Note: You are not allowed to use arrays to store the input.
\end{example}

\vspace{0.1 cm}
\begin{example}
Factors of a numbers are often required to know about the characteristics of a number.

In this problem, you'll print all prime factors of a given integer. (Prime numbers are the numbers which have exactly two factors i.e. 1 and itself).
You have to print all prime factors of a number in a new line in descending order. If a number is itself prime, print -1.
\end{example}

\vspace{0.1 cm}
\begin{example}
Write a program to implement a  rotation cipher as defined:

The program first reads three integers k1, k2 and k3 separated by white spaces. It then reads a characters from the NEXT line. Change the character according to the following rules:
(a) if it is a lower case character, it is rotated by k1 positions.  For example, if k1 is 3 then  `a' becomes `d', `b' becomes `e', ..., `x' becomes `a', `y' become `b',  `z' becomes `c'.
(b) if it is an upper case character, it is rotated by k2 positions. For example, if k2 is -3 then  `A' becomes `X', `B' becomes `Y', ..., `X' becomes `U', `Y' become `V',  `Z' becomes `W'.
(c) if it is a digit, it is rotated by k3 positions. For example, if k3 is 4 then  `3' becomes `7', `6' becomes `0', ..., `0' becomes `4', `5' become `9' and so on.
(d) Any other character remains the same.

The output is a single character obtained after above change.
\end{example}

\vspace{0.1 cm}
\begin{example}
Given two integer arrays (let's say A1 and A2), check if A2 is a contiguous subarray of A1. A2 is a contiguous subarray if all elements of A2 are also present in A1 in the same order and continuously.

For ex. [12,42,67] is a contiguous subarray of [1,62,12,42,67,96]
Whereas, [1,23,21] and [12,42,96] are not contiguous subarrays of [1,62,12,42,67,96]

Input: 
The first line contains the size N1 of first array.
Next line contains N1 space separated integers giving the contents of first array.
Next line contains the size N2 of second array.
Next line contains N2 space separated integers giving the contents of second array.

Output:
Either YES or NO (followed by a new line).

Variable Constraints:
The array sizes are smaller than 20.
Each array entry is an integer which fits an int data type.
\end{example}

\clearpage
\vspace{0.1 cm}
\begin{example}
You are given two integers n1 and n2 followed by two space separated strings str1 and str2 of length n1 and n2 respectively, each consisting of lowercase characters. The length of each of the strings is not more than 500.

Output the length of the initial segment of str1 which consists entirely of characters in str2.
\end{example}

\vspace{0.1 cm}
\begin{example}
You are given an array of `n' numbers. You have to find out whether the array is a SuperArray or not. An array is a SuperArray if it satisfies the following constraints. 

Every element A[i] of the array should occur A[i] times. For example if the array contains `2', then there should be exactly two occurrences of the number `2' in the array.
\end{example}

\vspace{0.1 cm}
\begin{example}
Find out whether or not a path exists through a given maze.
The maze is a 2D matrix where `.' denotes path and `X' denotes wall.
It starts at (0,0) and end at the bottom-right(both of which will always be `.')

Input:
Space separated integers m,n denoting size of matrix
Next m lines contain a string of n characters(composed of `.' and `X')

Input Constraints:
1<=m, n<=15

Output:
YES if path exists, NO otherwise
\end{example}

\vspace{0.1 cm}
\begin{example} 
In this exercise, you need to implement GCD. However, the challenge is that you are not allowed to return any values. So, the modified GCD function takes two pointers as follows:
void gcd(int *a, int *b)
It modifies the values such that when the function returns, a will contain the final answer. You need to use the function signature from the initial template.
\end{example}

\vspace{0.1 cm}
\begin{example} 
Write a program to find kth largest element of an array.

Full points will only be awarded if your solution is based on repeated applications of the Partition function (which was introduced for QuickSort). You do not have to sort the whole array, as this will fetch you half of the total points.

Any other solution e.g. solutions based on sorting, etc. will at most fetch half of total points.

Please see the provided template for hints.

Input will have two lines - 1st line will have an integer n denoting number of elements of the array and k; next line will contain n space separated integers denoting the elements of the array.

Output: you have to return the kth largest element of the array
\end{example}

\vspace{0.1 cm}
\begin{example}
The professor of PHY101A has decided to catch all cheating cases. Since you have already done that course, you decide to help him in this task by automating his work.

You are going to calculate the 'proximity' between any 2 documents by counting the longest common substring in the 2 documents.
For example,
- If one of the document is `ABA' and the other document is `BAB', the proximity is 2 since the longest common substring is `AB' (or `BA').
- If one of the document is `doc1' and the other document is `doc2', the proximity is 3 since the longest common substring is `doc'.

Input:
Two integers (`n1' and `n2') denoting the length of first and second document.
Content of first document (`n1' characters)
Content of second document (`n2' characters)

Output:
A single integer, the proximity between the documents
\end{example}

%%%%%%%%%%%%%%%%%%%%%%%

\clearpage
\subsection{Concrete Examples Illustrating Our Bug-Localization Results}

\vspace{0.4cm}
\begin{example}[Wrong \texttt{for} Loop]
\begin{lstlisting}
#include <stdio.h>
int rot(int [],int,int);
int main() {
    int n,d,i;
    scanf("%d\n",&n);
    int arr[n];
    for(i=0;i<n;i++) {
        scanf("%d ",&arr[i]); }
    scanf("\n%d",&d); 
    rot(arr,n,d); 
    return 0; } 

int rot(int arr[],int n,int d) {
    int j,k;
    for(j=d+1;j<n;j++) {       \\ suspiciousness score: 0.0006181474
        printf("%d ",arr[j]); }
    for(k=0;k<=d;k++) {     \\ suspiciousness score: 0.0006690205
        printf("%d ",arr[k]); }
    return 0; }
\end{lstlisting}

%\noindent\rlap{\rule{\dimexpr\textwidth-1cm}{1pt}}%

This program is supposed to right shift a given array of `n' numbers by a given number `d'.
To correctly implement this, the programmer needs to change the two \texttt{for} loops at lines $ 15 $ and $ 17 $ to 
\texttt{for(j=n-d;j<n;j++)} and \texttt{for(k=0;k<n-d;k++)}, respectively.
\Tool{} ranks these two lines as its second and third most suspicious buggy lines, respectively.
\end{example}

\vspace{0.3cm}
\begin{example}[Incorrect Input Reading and Output Formatting]
\begin{lstlisting}
#include <stdio.h>
int main(){ 
    int n,i;
    char c;    
    scanf("%d",&n);   \\ suspiciousness score: 0.0007697232
    for(i=0;i<n;i++) {
        scanf("%c",&c);
       if (c=='a'|| c=='e' || c=='i' || c=='o'|| c=='u') {
           printf("Special");
           printf("\n%d",i);   \\ suspiciousness score: 0.00045288168
           break; } } 
    if(i==n) 
    printf("Normal");
    return 0; }
\end{lstlisting}

This program is supposed to print `Special' if the given input string contains a vowel, otherwise `Normal'.
The input format is an integer `n' and a string `s' of length n, separated by a newline character.
However, the \texttt{scanf} function in line $7$ reads the newline character following `n' as the first character of the string. 
Therefore, if the input string is a vowel, that will not be read and the program will print the wrong output `Normal'.
One way to fix it is to append the newline character after the ``\%d'' format specifier in the \texttt{scanf} function call of line $4$.
Also, there is an additional print statement at line $10$ which prints spuriously, causing an output mismatch.
\Tool{} ranks these two as its third and fourth most suspicious buggy lines, respectively.
\end{example}

\clearpage

%\vspace{0.3cm}
\begin{example}[Wrong Condition]
\begin{lstlisting}
#include <stdio.h>
int main() {
    int n,i;
    char a[100];
    char b;
    int flag=0;
    scanf ("%d",&n);
    for (i=0;i<n;i=i+1) {
         scanf ("%c",&b);
         if((b=='a')||(b='e')||(b='i')||(b=='o')||(b=='u'))  \\ suspiciousness score: 0.0015987115
         flag=1; }
     if(flag==1) {
         printf("Special"); }
     else  {
         printf ("Normal"); }
    return 0; }
\end{lstlisting}

The program shown above solves the same problem as the program last discussed. It twice uses the assignment operator instead of the comparison operator in line $10$ which causes the bug. \Tool{} localizes it in its top prediction.

\noindent\rlap{\rule{\dimexpr\textwidth-1cm}{1pt}}%

\begin{lstlisting}
#include<stdio.h>
int main() {
    int a,b,c;
    scanf("%d%d%d",&a,&b,&c); 0.00079781574
    if (a+b>c) { 0.00025660603
        if (a*a+b*b==c*c){printf("RIGHT");}
        else if(a*a+b*b<c*c||a*a>b*b+c*c||a*a+c*c<b*b){     \\ suspiciousness score: 0.0004023624
			printf("OBTUSE"          );} 
        else if(a*a+b*b>c*c||a*a<b*b+c*c||a*a+c*c>b*b){     \\ suspiciousness score: 0.00045172646
			printf("ACUTE");} } 
    else if(a+b==c) {printf("INVALID");} 
    return 0; }
\end{lstlisting}

The above program is supposed to check and print if a triangle is invalid, acute, right or obtuse, given the length of its three sides.
However, the conditions used in lines $ 7 $ and $ 9 $ are buggy. To fix the program, they should be replaced by lines (1)
\texttt{else if(a+b>c \&\& a+c>b \&\& b+c>a) \{} and (2)
\texttt{else \{}, respectively.
\Tool{} ranks them as its third and fourth most suspicious buggy lines, respectively.
\end{example}

\clearpage

%\vspace{0.3cm}
\begin{example}[Insufficient memory allocation]

\begin{lstlisting}
#include <stdio.h>
int in(int k,int n,int l[100]){
    int i;
    for(i=0;i<n;i++){
        if(l[i]==k){
            return 1; } }
    return 0; }
int main(){
    int n;
    int ip[100];
    int u[100];  \\ suspiciousness score: 0.0012746735
    scanf("%d",&n);
    int i;
    for(i=0;i<n;i++){
        scanf("%d",&ip[i]); }
    int k=1,count=0;
    i=0;
    while(!in(k,n,u)){
        u[i]=k;
        k=ip[k-1];
        i+=1;
        count+=1; }
    printf("%d ",count);
    for(i=0;i>=0;i++){
        if(u[i]==k){
            printf("%d",count-i);
            break; } }
    return 0; }
\end{lstlisting}

The program shown declares an array of fixed-size in line $ 11 $ which fail on tests containing larger inputs. \Tool{} localizes the buggy statement it in its top prediction.
Note that the other fixed-sized array declared in line $ 10 $ is not considered buggy as it does not cause any available test to fail.
\end{example}

\vspace{0.3cm}
\begin{example}[Type Narrowing]
\begin{lstlisting}
#include <stdio.h>
int main(){
    int x1,y1,x2,y2;
    float slope;
    scanf("%d%d%d%d",&x1,&y1,&x2,&y2);
    if(x1==x2)  {
        printf("inf");
        return 0; }
    else {
        slope==(y2-y1)/(x2-x1);   \\ suspiciousness score: 0.0028934027
        printf("%.2f\n", slope); }
    return 0; }
\end{lstlisting}

This program calculates the slope of a line specified by two points whose coordinates are given as four integers $ (x_1, y_1) $ and $ (x_2, y_2) $. When calculating slope in line $10$, the division of integers returns integer value and not a floating point value. This is known as narrowing of types and can be fixed with type-casting any of the variable or expression in the RHS of the assignment as float before the division operation. The buggy line also mistakenly uses a comparison operator instead of the assignment operator.
\Tool{} localizes the buggy line in its top prediction.
\end{example}

\clearpage

%\vspace{0.3cm}
\begin{example}[Wrong Assignment]
\begin{lstlisting}
#include <stdio.h>
#include<string.h>
int main() {
    int i,j,c;
    char str1[10],str2[10];
    scanf("%s %s",str1,str2);
     c=strlen(str2);
    for(i=0;str1[i]!='\0';i++) {
      str1[i]=str1[i]+str2[i%c]-'a'+1; }  \\ suspiciousness score: 0.0011707128
  printf("%s",str1);
    return 0; }
\end{lstlisting}

The program shown above is supposed to shift a string by another pattern string~(both given as input) and print the result.
However, the RHS expression of the assignment at line $ 9 $ is buggy. The correct RHS expression is:
\texttt{str1[i]=(str1[i]+str2[i\%c]-\textquotesingle a\textquotesingle-\textquotesingle a\textquotesingle+1)\%26+\textquotesingle a\textquotesingle;}. 
\Tool{} localizes the buggy line in its top prediction.

\noindent\rlap{\rule{\dimexpr\textwidth-1cm}{1pt}}%

\begin{lstlisting}
#include <stdio.h>
int main(){
   int a,b,i,n,m;
   scanf("%d%d%d",&a,&b,&m);
   n=1;
   for (i=1;i<=b;i=i+1)
   n=n*a;    \\ suspiciousness score: 0.007239882
   printf("%d",n%m);
  return 0; }
\end{lstlisting}

The program shown above implements $ a^b \bmod m $. However, the RHS expression in line $ 7 $ does not implement this logic correctly. The fix for this line is: \texttt{n=(n*a)\%m;}.
\Tool{} localizes the buggy line in its top prediction.
\end{example}

\vspace{0.3cm}
\begin{example}[Missing Code]
\begin{lstlisting}
#include <stdio.h>
int main(){
    int n,max=0,sum,i,j=0;
    scanf("%d/n",&n);
    char s[n],ch;
    ch=getchar();
    for(i=0;i<n;i++)
    {ch=getchar();
    s[i]=ch;}
    for(i=0;i<n;i++)
    { sum=0;    \\ suspiciousness score: 0.0013130781
        while(s[i]==s[i+j])
        {sum++;
        j++; }
        if(max<=sum)
        max=sum; }
    printf("%d",max);  
    return 0; }
\end{lstlisting}

This program is written for finding the longest contiguous streak of a character in a given string.
To correctly implement this, the programmer needs to insert \texttt{j=0;} at line $11$. \Tool{} localizes this bug in its top prediction.
\end{example}

\end{document}